\documentclass[a4paper]{aa}
\def \th {\thinspace}

\def \src {X\thinspace 1624-490}
\def \degmark{^\circ}

\def\approxgt{\mathrel{\hbox{\rlap{\lower.55ex \hbox {$\sim$}}
\kern-.3em \raise.4ex \hbox{$>$}}}}
\def\approxlt{\mathrel{\hbox{\rlap{\lower.55ex \hbox {$\sim$}}
\kern-.3em \raise.4ex \hbox{$<$}}}}
\def \th {\thinspace }
\def \ref {\reference{}}

\def \sun {\hbox {$\odot$}}
\def \degmark{^\circ}

\usepackage{graphicx}
\begin{document}

%\thesaurus{6(13.25.5;  % X-rays: stars,
%             08.09.2: X\th 1624-490;  % stars: individual: \src,
%             08.14.1;  % stars: neutron,
%             08.02.1;  % binaries: close,
%             02.01.2)} % accretion, accretion disks

\title{Neutron star blackbody contraction during flaring in X\th 1624-490}

\author{M. Ba\l uci\'nska-Church\inst{1,2}
\and R. Barnard\inst{1}
\and M. J. Church\inst{1,2}
\and A. P. Smale\inst{3}}

\offprints{M. Ba\l uci\'nska-Church}

\institute{School of Physics and Astronomy, University of Birmingham,
           Birmingham B15 2TT\\
           email: mbc@star.sr.bham.ac.uk
\and
          Institute of Astronomy, Jagiellonian University, ul. Orla 171, 30-244 Cracow, Poland
\and
          Laboratory for High Energy Astrophysics, Code 662,
          NASA/Goddard Space Flight Center,\\
          Greenbelt, MD~20771, USA.\\
          Also, Universities Space Research Association
                    email: alan@osiris.gsfc.nasa.gov}
                    
%\thanks{}

\date{Received 12 January 2001; Accepted 16 August 2001}

\authorrunning{Ba\l uci\'nska-Church et al.}

\titlerunning{Flaring in X\th 1624-490}

\abstract{We present results of an extensive investigation of the physical
changes taking place in the emission regions of the LMXB X\th 1624-490 during
strong flaring in observations made using {\it RXTE} in 1997 and 1999.
Based on the detailed light curve, 
we propose that the flaring consists of a superposition of X-ray bursts.
It is shown that major changes take place in the blackbody emission
component, the temperature $kT_{\rm {BB}}$ increasing to $\sim $2.2 keV
in flaring. Remarkably, the blackbody area decreases by a factor of
$\sim $5 in flaring.
During flare evolution, the blackbody luminosity remains approximately
constant, constituting a previously unknown Eddington limiting effect
which we propose is due to radiation pressure of the blackbody
as $kT_{\rm BB}$ increases affecting the inner disk or accretion flow 
resulting in a decreased emitting area on the star. We argue that the
large decrease in area cannot be explained in terms of modification
of the blackbody spectrum by electron scattering in the
atmosphere of the neutron star. The height of
the emitting region on the non-flaring neutron star is shown to agree with the
height of the inner radiatively-supported accretion disk 
as found for sources in the {\it ASCA} survey of
LMXB of Church \& Ba\l uci\'nska-Church (2001). The decrease in height
during flaring is discussed in
terms of possible models, including radial accretion flow onto the
stellar surface and the theory of accretion flow spreading on the 
neutron star surface of Inogamov \& Sunyaev (1999). We demonstrate
that the intensity of the broad iron line at 6.4 keV is strongly correlated
with the luminosity of the blackbody emission from the neutron star,
and discuss the probable origin of this line in the ADC. Finally,
possible reasons for non-detection of a reflection component in this source, 
and LMXB in general, are discussed.
}

\maketitle

\section{Introduction}
\src\ is one of the most unusual members of the class of dipping Low Mass X-ray Binary
(LMXB) sources exhibiting periodic dips in X-ray intensity. It is generally accepted
that dipping is due to absorption in the bulge in the outer accretion disk where the
accretion flow from the companion impacts (White \& Swank 1982). \src\ has the longest
orbital period of the dipping sources at $\sim $21 hr. Dipping is deep, $\sim$ 75\% in 
the band 1--10 keV, and the source also exhibits strong flaring in which the X-ray flux 
can increase by 30\% over timescales of a few thousand seconds (Church \& 
Ba\l uci\'nska-Church 1995). Flaring has been known in LMXB for many years since it was 
recognized that \hbox {Sco\th X-1} had quiescent states and active states in which 
strong X-ray flaring took place (Canizares et al. 1975). White et al.
(1985), using the {\it Exosat} GSPC, showed that the continuum between 2--25 keV 
could be modelled by 
a blackbody plus unsaturated Comptonization spectrum  and that the flaring is caused 
by increases in the blackbody luminosity from 10\% to 40\% of the total. The blackbody 
temperature increased significantly in flaring together with some increase in emitting area. 
Hasinger et al. (1989) realized that Sco\th X-1 was a Z-track source, 
the source having different spectral states forming a Z-track on a hardness-intensity diagram 
with associated changes in QPO behaviour. It was also realized that
several other bright sources exhibited similar flaring to that in Sco\th X-1.
However, since then, little work has been carried out on flaring in
LMXB, physical understanding of the flaring phenomenon is limited,
and the transition between bursting in less bright LMXB and flaring 
in brighter sources is not understood.

Investigation of X-ray dipping has however, been more extensive. The depth, duration and 
spectral evolution in dipping vary considerably from source to source. Moreover, spectral 
analysis of dipping sources provides information not available in
non-dipping sources. For 
example, spectral models are more strongly constrained by having to fit non-dip and dip 
data, thus showing clearly the nature of the emission. It has been shown that the dipping 
LMXB sources are well described by assuming point-like blackbody emission from the
neutron star plus Comptonized emission from an extended Accretion Disk Corona (ADC) 
(e.g. Church \& Ba\l uci\'nska-Church 1995; Church 2001). In addition, the dip ingress and 
egress times can be used to obtain the sizes of the extended emission region if the
absorber has larger angular extent that the source regions, as proven by dipping being 
100\% deep at any energy, which is often the case. This provides a radius for the ADC, 
$r_{\rm ADC}$, typically 50,000 km or 15\% of the accretion disk radius, although in
X\th 1624-490, $r_{\rm ADC}$ is $\rm {5.3\times 10^{10}}$ cm or 50\%
of the disk radius 
(Church 2001). There is also evidence that the ADC is ``thin'', having height-to-radius ratio
of $\sim $10\% (Smale et al. 2001). This shows that the ADC is a very extended, flat 
region above the accretion disk, contrary to theoretical suggestions often made that 
the Comptonizing region may be localized to the neighbourhood of the
neutron star.

Recent results of an {\it ASCA} and {\it BeppoSAX} survey of LMXB 
(Church \& Ba\l uci\'nska-Church 2001)
reveal a surprising result for the area of the neutron star emitting
blackbody radiation for sources of different luminosities. When it is assumed that 
the emission region is an equatorial strip of half-height {\it h}, it
is found that there is good agreement between {\it h} and {\it H}, the
half-height of the inner disk, assumed to be radiatively supported. It
is not clear at the moment whether there is a direct causal link between these 
parameters, for example, if material flowed radially between the inner disk and the 
star, or whether there is an indirect link such that {\it H} is a measure of
the total source luminosity $L_{\rm {Tot}}$ and some physical process depending
on $L_{\rm {Tot}}$ determines the extent of the emitting area. 
In particular, Inogamov \& Sunyaev (1999) have 
shown that accretion flow regarded as a fluid meeting the neutron star at the equator 
is expected to spread vertically up the star to a height dependent
on $L_{\rm {Tot}}$. A preliminary comaparison of this model with the
survey results (Church et al. 2001) shows that observed blackbody
luminosities are $\sim$3 times larger than implied by the theory,
suggesting that radial flow may be the dominant factor determining the
blackbody luminosity. The results of the present work will be
viewed (Sects. 3 and 4) in terms of the {\it h = H} relation.

\src\ was previously observed in a long {\it Exosat} observation of 220 ks and with {\it Ginga}
(Jones \& Watson 1989). The {\it Exosat} ME data revealed an apparently
stable lower level of dipping supporting the presence of two emission components, one of
which was totally absorbed in deep dipping. Church \& Ba\l
uci\'nska-Church (1995) showed that the light curve at higher energies ($>$ 5 keV)
was dominated by flaring which can strongly modify the spectrum. Thus if
flaring occurs simultaneously with dipping, the investigation of spectral
evolution in dipping will be difficult.
By selecting sections of non-dip and dip data
without apparent flaring, the above two-component model was found to fit the data well
showing that in deep dipping the blackbody component was totally absorbed, and that
the Comptonized component was relatively little absorbed. However, with only one
non-dip and one deep dip spectrum it was not possible to determine
the extent of absorption of the Comptonized component. The Galactic
column density of \src\ is very high ($\sim \rm {8\times 10^{22}}$ atom
cm$^{-2}$) so that a dust scattered halo of the source is expected, and
Angelini et al. (1997) demonstrated an excess of the surface brightness
above the point spread function in {\it ASCA} GIS. 

The 60 ks observation
with {\it BeppoSAX} of 1999, August 11 was almost free of flaring and
fitting the radial intensity profile of the MECS instruments in four
energy bands between 2.5 and 6.5 keV allowed a value of the optical depth to dust scattering
of 2.4$\pm$0.4 to be determined (Ba\l uci\'nska-Church et al. 2000).
Moreover, the very broad band of {\it BeppoSAX} allowed the
Comptonization cut-off energy to be determined for the first time,
this having the value of $\rm {12^{+14}_{-5}}$ keV. Spectral evolution
in dipping could be well-described in terms of an emission model
consisting of point-like blackbody emission from the neutron star
plus extended Comptonized emission from the ADC. It was shown that
using a disk blackbody model instead of a simple blackbody resulted in
an unphysical inner radius of 0.4 km, so this possibility could be
discounted. Similar unphysical results were obtained for this model 
for the majority of the LMXB in the {\it ASCA} survey allowing us to
associate the blackbody emission with the neutron star in LMXB in
general. Additionally, the extended size of the ADC and high optical
depth imply that all of the thermal emission from the disk will be
reprocessed in the ADC. Thus, we do not expect any contribution to the
observed thermal emission from the disk.

Several {\it RXTE} observations have been made
by the authors including observations on 1997, January 6--7,
1997, May 4, 6, 11 which failed to capture dipping. A 4-day observation with
{\it RXTE} was made in 1999, September spanning 4 orbital cycles with
deep dipping and extensive flaring. Results of analysis of the
spectral evolution in dipping are presented in Paper I (Smale et al.
2001). The orbital period has not been refined since the {\it Exosat}
determination of 21$\pm $2 hr (Watson et al. 1985). In Paper I,
an orbital ephemeris and improved period of 20.8778$\pm $0.0003 hr
are presented based on 4.5 yr of {\it RXTE} ASM data. In addition, a search
failed to detect QPO at frequencies from Hz to kHz, which may be related
to the high luminosity of the source. The much higher count rate and
observation duration in the 1999, September observation with {\it RXTE}
revealed several new features including the dominance of electron scattering
in the early stages of dipping. The main deep dips were surrounded by
shallow ``shoulders of dipping'' in which a vertical shift
of the spectrum relative to the non-dip spectrum was seen
between 2.5 and 25 keV. This demonstrated that electron scattering was dominant
and photoelectric absorption was not detectable
indicating the highly ionized state of the outer absorber produced by the
high luminosity of X\th 1624-490. A broad iron line at $\sim $6.4 keV was 
first detected by Asai et al. (2000) using {\it ASCA}.
Spectral evolution in dipping could be well described
using the above two-component emission model having the basic form:
{\sc ag$\ast$(ab$\ast$ bb + pcf$\ast$ (cpl + gau))}, where {\sc ag} represents Galactic
absorption, {\sc ab} additional absorption of the point-like blackbody in
dipping, {\sc pcf} the progressive covering of the extended Comptonized
emission (described as a cut-off power law) and {\sc gau} the Gaussian line.
Good fits were obtained by including the 6.4 keV line in the same covering
term as the Comptonized emission, showing that the line probably originates in
the ADC. However, to the basic form were added complicating terms for the effects of
both electron scattering and dust scattering (Smale et al. 2001).

In the present paper (Paper II), we present a study of flaring in \src\ based on 
the observations of 1997, January and 1999, September during both of which strong 
flaring took place. Although flaring in Sco\th X-1 and similar sources has been 
known for many years, in which clear indications of increasing blackbody temperature 
were found, there has been no recent observational work aimed at 
understanding the physical changes taking place, or the differences between bursting and
flaring sources. In addition, flaring is clearly related to the flaring branch of the 
Z-track class of LMXB. \src\ offers a unique opportunity for investigating these problems
as it is both a dipping and a flaring source.

\section{Observations}
\label{sec:observations}

\src\ was observed using {\it RXTE} (Bradt, Rothschild, \& Swank, 1993)
from 1999, September 27 18:22 UT -- \hbox {October} 1 13:11 UT, 
for a total on-source exposure of 212 ks. In addition to results from this main
observation spanning 4 orbital cycles of 20.87 hr, we present results from our observation of 
1997, January 6 11:14 UT -- January 7 3:28 UT. The new ephemeris (Paper I)
shows that this observation, intended to capture dipping, spanned
phases 0.11--0.89 where phase 0 is dip centre, exactly missing deep
dipping. However, there was a period
of strong flaring and both observations are of interest because of
the change in source luminosity between the observations, the 1997
observation having a luminosity (1--30 keV) of $1.06 \times 10^{38}$ erg s$^{-1}$,
compared with $1.35 \times 10^{38}$ erg s$^{-1}$ in the 1999 observation, approaching
closer to the Eddington limit for a $\rm {1.4\,M_{\sun}}$ object. 
It will be seen that this difference apparently leads to significant
differences in flare development (Sect. 3). For the analysis presented here of
relatively slow flare development and decay, data from the PCA instrument 
in the Standard 2 mode were used having a time resolution of 16 sec. 
The PCA consists of five Xe proportional counter units (PCUs) numbered 0 to 4, with 
a combined total effective area of about 6500 cm$^2$ (Jahoda et al. 1996). During the 
September 1999 observation only PCUs 0 and 2 were reliably on throughout
the observation, whereas in 1997 all PCUs could be used. To allow direct comparison of the
lightcurves and spectra, results are presented here for both observations using PCUs 0 
and 2, the high count rate permitting this. Data were also obtained with the HEXTE
phoswich detectors, which are sensitive over the energy range
15--250~keV (Rothschild et al. 1998). However, because of the 
relatively low break energy of the Comptonized spectrum at 12 keV
(Paper I) producing a relatively low count rate in HEXTE, and because spectral 
evolution during flaring takes place 
primarily in the band 1--20 ~keV, the HEXTE data provided little
useful additional information.
\begin{figure}
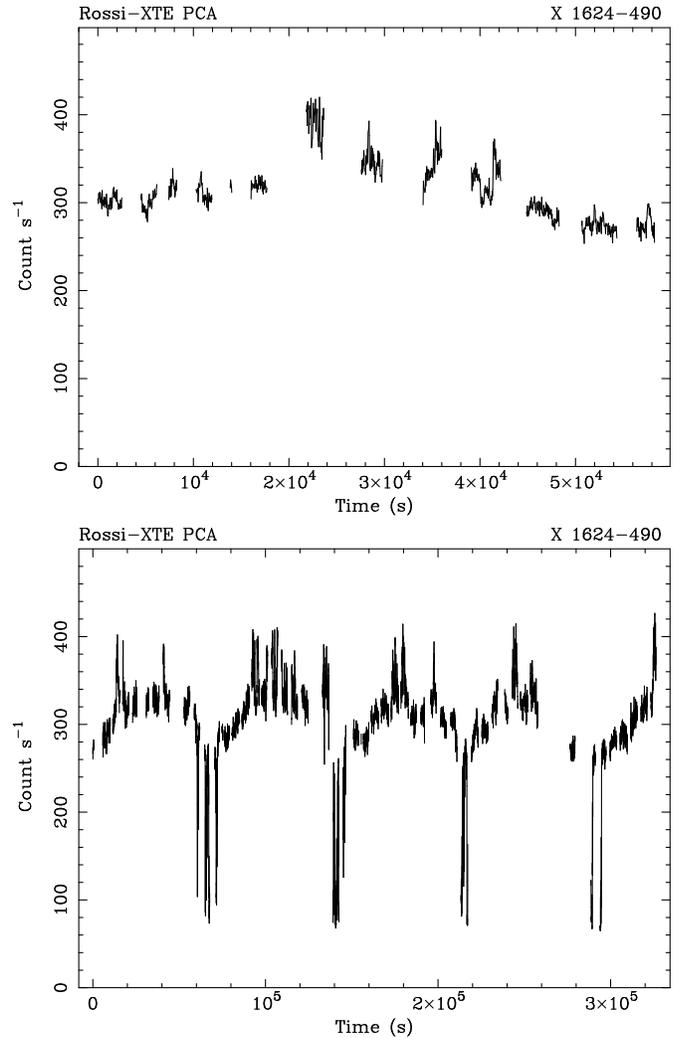

\includegraphics[width=68mm,angle=270]{f1a}     %orig lc_jan97_pcu02
\vskip 1 mm
\includegraphics[width=68mm,angle=270]{f1b}     % orig lc_sep99_pcu02
\caption{PCA light curve of X\th 1624-490 with 32s binning, upper curve: 
the 1997, January observation in the band 1.9--25 keV; lower curve: 
the 1999, September observation in the band 2.0--25 keV}
\end{figure}
\begin{figure}[!h]                                             %Fig. 2
\includegraphics[width=63mm,angle=270]{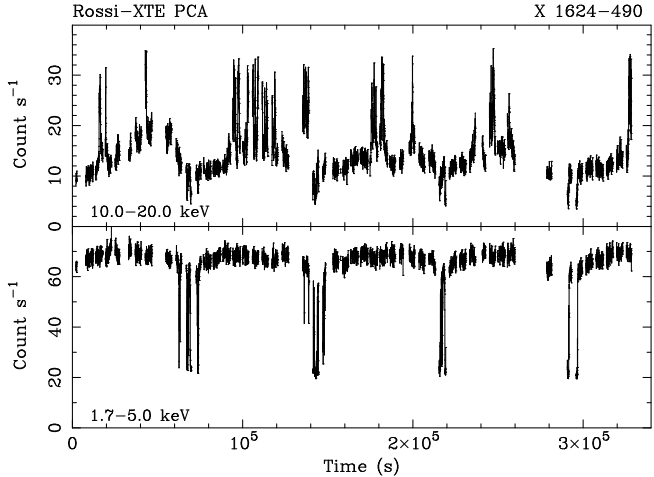}      % orig both32
\caption{PCA light curves of the 1999 observation of X\th 1624-490
in two energy bands: 1.7--5.0 and 10.0--20.0 with 32 s binning}
\end{figure}

Data analysis was carried out using the {\sl RXTE} standard analysis
software, FTOOLS 5.0.1. Background subtraction of the PCA data was
performed using the ``skyvle/skyactiv 20000131'' models generated
by the {\sl RXTE} PCA team. The quality of the background subtraction
was checked firstly by comparing the source and background
spectra and light curves at high energies (50--100 keV) where the source
no longer contributes detectable events; and also using the
same models to background-subtract the data obtained during slews to
and from the source. We conclude that our background subtractions in
the 2--20 keV energy range are accurate to a fraction of a count per
second. The quality of both the background subtraction and
the response matrices used was checked by including a careful analysis of Crab
calibration observations on 1999, September 26, October 13, November 8,
and November 23, before and after our 1999 observation of X\th 1624-490
(see Paper I). Systematic errors of 1\% were added to the spectra.

\section{Results}

\subsection {The X-ray lightcurves}

Figure 1a and 1b show light curves of the two observations in the
total PCA band. As the 1997 observation was made during PCA gain epoch 3,
and the 1999 observation during epoch 4, we used
appropriate PCA channel ranges to give very similar energy ranges in
the two light curves: 1.9--25 keV and 2.0--25 keV for 1997 and 1999,
respectively. Figure 2 shows the 1999 observation in 
two energy bands: a low energy band (1.7--5.0 keV) in which dipping is 
predominant and a high energy band (10.0--20.0 keV) in which flaring dominates.

In the 1997 lightcurve, a period of strong flaring is followed almost certainly by 
a shoulder of dipping, i.e. the period of shallow dipping that often precedes and 
follows the main deep dipping. In Paper I, we showed that this consists of a 
reduction in intensity of the extended Comptonized emission by electron scattering 
in the outer layers of the absorber before the neutron star is
overlapped by absorber and deep dipping commences. During the
shoulder, the blackbody emission component was shown to have
zero column density additional to the Galactic contribution; however,
it is important to realize that this does not represent the quiescent, i.e.
non-flaring, non-dipping emission which has a somewhat higher intensity. 
The 1999 lightcurves in Fig. 2 demonstrate vividly that while dipping is 
essentially a low-energy phenomenon, flaring dominates the light curves 
at high energies, and can only, in fact, be detected above
$\sim $~6 keV. The reason for this can be understood in
terms of the non-flare blackbody temperature
$kT_{\rm {BB}}$ $\sim $~1.3 keV which increases to 
$\sim $~2.2 keV in flaring, so that the spectra cross
over at about 6 keV, resulting in a deficit below
this energy and an excess above, compared with the non-flare spectrum.
Close inspection of the 1999 light curve indicates that each flare
is highly structured consisting of many shorter events, and we propose
here that flaring consists of a superposition of X-ray bursts
occurring with much reduced separation compared with
burst sources, due to the mass accretion rate being more than 10 times higher
(at a luminosity of $\rm {\sim 10^{38}}$ erg s$^{-1}$). 

\subsection {Spectral evolution in flaring}

\subsubsection {Selection of flare spectra}
\begin{figure*}[!ht]
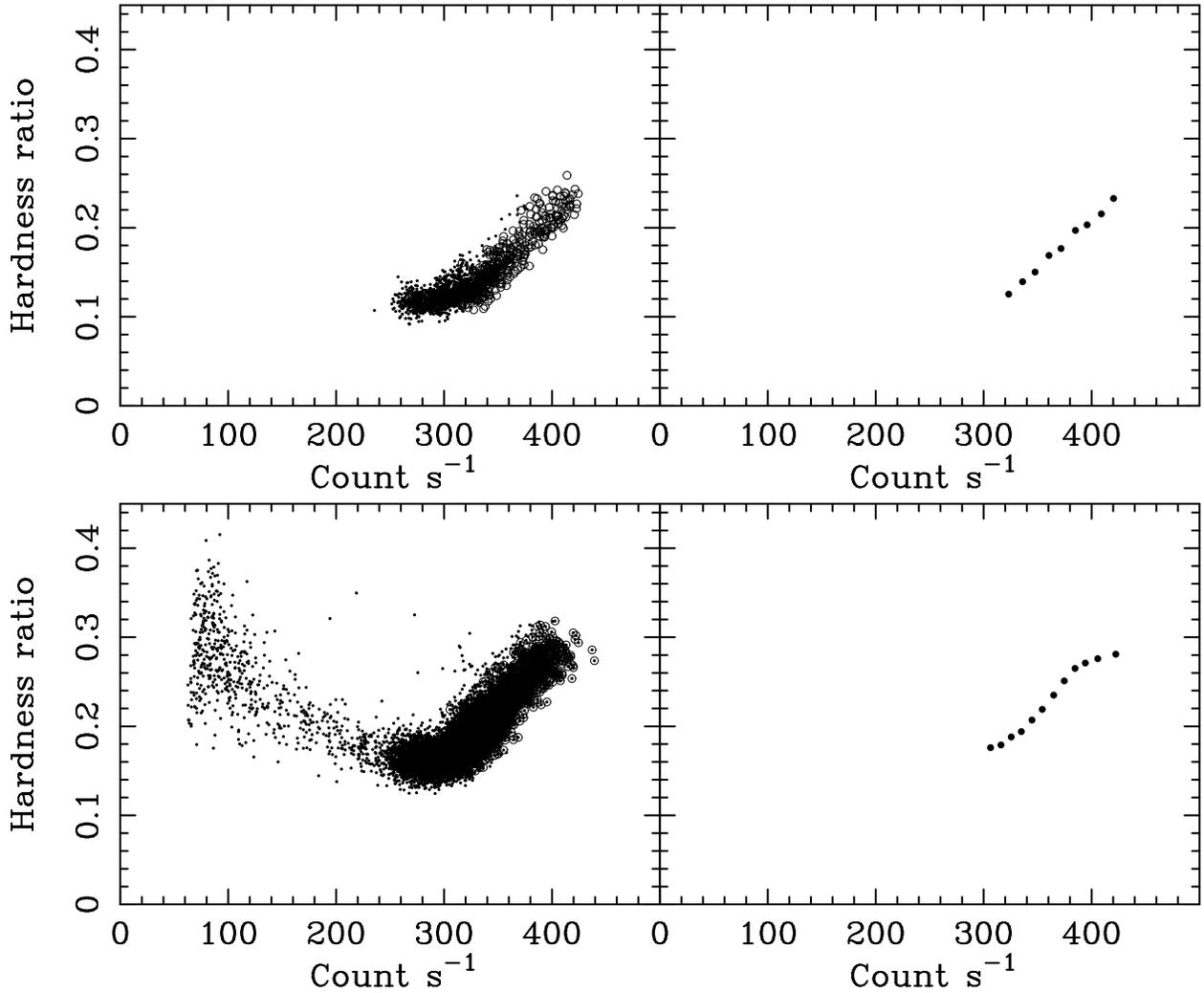
                                            %Fig. 3
\begin{center}
\includegraphics[width=68mm,angle=270]{f3b}  % - was [hr2_int_jan97_02]
\vskip 1 mm
\includegraphics[width=68mm,angle=270]{f3a}  % - was [hr2_int_sep99_02]
\caption{Flare evolution displayed on hardness-intensity diagrams; 
upper panel left: the 1997 observation; lower panel left:
the 1999 observation. The dots show data that was excluded during
selection of flare data, i.e. data in dipping, the shoulders of dipping, 
or flaring during the shoulders; open circles shows data accepted for
flare analysis. The right hand panels in each case show the corresponding
plots for spectra selected for analysis, allowing these to be compared
with the total data (see text)}
\end{center}
\end{figure*}
Flare evolution is firstly displayed on X-ray hardness-intensity
diagrams for both the 1997 and 1999 observations in Fig. 3
(left panels). Hardness is defined as the ratio of intensities in the
bands 9.8--24.9 keV and 4.9--9.8 keV, and intensity is in the total
band 1.9--24.9 keV. In the 1997 data (upper panel), strong flaring but only 
shallow dipping took place, whereas in the 1999 data (lower panel), there 
was both strong flaring and deep dipping. In this case, dipping begins with  
intensity decreases of $\sim $15\% in an energy-independent
way (horizontal part of Fig. 3) corresponding to a shoulder of dipping,
and consistent with electron scattering of the non-dip spectrum 
in highly ionized outer layers of the absorber (Paper I). In deeper
dipping, spectral evolution is complex, showing first a strong hardening
as Comptonized emission is absorbed, then a softening in deepest dipping 
(at $<$ 80 c s$^{-1}$) as the blackbody emission becomes totally removed 
leaving the residual spectrum harder.  

During flare evolution, systematic changes clearly take place in hardness-intensity
very similar to the flaring branch of Z-track sources displayed on similar diagrams.
Flare spectra selected for spectral analysis are also shown as
average points for each spectrum in Fig. 3 (right panels). It was not
initially clear that simply selecting flare
spectra on the basis of intensity would be adequate to allow
understanding of spectral evolution during flaring. Preliminary tests
with this and other sources displaying track movement revealed that
selection on intensity alone may produce scatter in the positions of the selected
spectra about a line drawn through the centre of the track on the
hardness-intensity plot. Such scatter leads to changes in spectral parameters not due to
the development of flaring particularly in the Comptonized emission,
and this complicates interpretation. Consequently it was
decided to select spectra in such a way that the points lay 
along the centre of the track on the hardness-intensity diagram that
represents flare evolution. This was done for the 1997 data by
selecting data on the basis of both hardness ratio and intensity, defining essentially
a box for each spectrum within Fig. 3. It can be seen that the spectra
produced well represent the general track movement. In the case
of the 1999 observation, spectra were first selected on the basis of
intensity alone, and when these data were mapped onto the
hardness-intensity diagram, it was found that these points moved systematically along the flare
evolution track, obviating the need for selection on both hardness and intensity. This
was a result of the much improved total count in spectra of the long observation.
Comparison of results from this `track-centre' technique with initial
tests showed that the behaviour during flare development becomes systematic.

For the 1997 observation, 9 flare spectra were selected using good-time intervals 
obtained from a hardness-intensity diagram made in the same energy bands as in Fig. 3 
but using all 5 PCUs to reduce the errors on every point. Only the first
3 flares were used because the 4th flare is in the shoulder of dipping.
For the 1999 observation, 12 flare spectra were selected in the same way
avoiding dipping and shoulder-of-dipping data. The results of fitting spectra selected 
in this way are discussed in the following section. Finally, we
investigated whether it is adequate to add data from different flares
by the procedure of intensity selection, and analysed data from single
flares including the strong flare at $\sim $180 ks in the 1999 data. 
(Fig. 1b). Results showed that this flare followed the same track on
the hardness-intensity diagram as the total 1999 data, and spectral
analysis results were also very similar. Consequently there was no need
to restrict analysis to individual datasets.

\subsubsection{Spectral evolution during flaring}

%\tabcolsep 3.0
\begin{table}[!ht]                      %Table 1
\caption{Ability of various spectral models to fit the 1999 flare
spectra simultaneously. The models are variations on the two-component
{\sc bb + cpl} model, except for the first, in which a second blackbody
is added to attempt to model flaring}
\begin{flushleft}
\begin{tabular}{llrr}
\hline\noalign{\smallskip}
$\;\;\;$&Model & $\chi^{\rm 2}$/dof&$\;\;\;$\\
\noalign{\smallskip\hrule\smallskip}
&Additional blackbody & 1038/589\\
&Constant cut-off power law & 813/568\\
&Constant blackbody & 1493/568\\
&Constant blackbody radius & 904/557\\
&Constant blackbody temperature & 1283/557\\
&Constant power law index & 445/557\\
&Constant power law normalization& 370/557\\
&Free fitting & 354/546\\
\noalign{\smallskip}
\hline
\end{tabular}
\end{flushleft}
\end{table}

In Paper I, it was shown that non-dip, non-flaring emission can
be well-described by a two-component model consisting of point-like
blackbody emission from the surface of the neutron star plus extended
Comptonized emission from the accretion disk corona represented
by a cut-off power law, plus a line at 6.4 keV. Dust scattering is significant
in this source, since photons from non-dip emission are scattered into
the beam in dipping due to the time-delay in scattering.
As the amount lost from the beam depends on the dip intensity; there is
an excess due to the difference of these terms which makes a
(non-dominant) contribution to the residual count rate in deep dipping.
Outside dipping, the two terms balance and no effect is expected, and the
spectrum is well-modelled by the two continuum components above.
The cross-section for dust scattering has a strong energy dependence
$\sim E^{\rm {-2}}$ therefore the limited effects that are seen are
at low energies and flaring at energies above 6 keV should not be influenced by
dust scattering. In Paper I
we exhaustively tested a range of one-component and two-component
models using non-dip and several dip spectra, which showed that only
the above model was capable of fitting all of these spectra. Consequently in
the
present work we use that model to investigate spectral evolution
during flaring, i.e. the model {\sc ag$\ast$(bb + cpl + gau)} where {\sc ag}
represents Galactic absorption, {\sc bb} is the neutron star blackbody,
{\sc cpl} is the cut-off power law representation of Comptonization
in an extended ADC and {\sc gau} is the Gaussian line. 

This model was firstly used to test various possibilities for flaring
evolution, such as the Comptonized emission remaining constant, by
fitting the 12 spectra from 1999 simultaneously, and
results are shown in Table 1. The probability of any model being
correct was vanishingly small except for the last three,
including attempting to model flaring by an {\it additional} blackbody
component. The last model with all parameters free was more probable
and acceptable at
$>$99\% probability, and so this model was used in fitting the 1997
and 1999 observations.

Spectral fitting was carried out by simultaneously fitting all flare
spectra in each of the observations using the two-component continuum
model described above together with a line at 6.4 keV with width $\sigma $
fixed at 0.4 keV (as in Paper I) to prevent the tendency of a broad line to
absorb neighbouring continuum. Initially, the Galactic column density was 
free, however, this is not well-determined in {\it RXTE}
and with free-fitting, values tended to become substantially higher
(50\%) than the well-determined value from {\it BeppoSAX} of
$\rm {8.6\pm 1.0\times 10^{22}}$ atom $\rm {cm^{-2}}$ (Ba\l
uci\'nska-Church et al. 2000). Consequently, this value was fixed in the fitting. Using {\it
BeppoSAX} with a spectrum extending to higher than 100 keV, it was also
possible to obtain a good measurement of the Comptonization cut-off
energy,
while in the {\it RXTE} PCA with more restricted energy band it is
difficult to get reliable values, particularly of possible changes in
this parameter during flaring. We show below that the most dramatic
changes in flaring take place in the
blackbody component; nevertheless, we tested
the effect of different approaches to the cut-off energy as follows.
Firstly, we allowed this parameter to be entirely free, or free but
chained between individual flare spectra in each observation.
Secondly, we fixed the value for all spectra at the best-fit value
obtained from {\it BeppoSAX} of 11.8 keV. Finally, we fixed the value
at various values between 11.8 keV and 100 keV. The quality of fit
deteriorated as the value was increased, and maximum values
of $E_{\rm {CO}}$ of 12 keV (1997) and 21 keV (1999) were found,
after which $\chi^2$/dof became unacceptable.
With the cut-off energy free but chained between spectra,
values of $\sim $5 keV (1997) and $\sim $9 keV (1999) were obtained.
However, the results presented below were found not to be sensitive
to $E_{\rm {CO}}$, particularly the blackbody temperature and radius,
and the systematic changes in these during flare development;
consequently it was decided to adopt the approach of fixing the
cut-off energy at the {\it BeppoSAX} value.
Spectral fitting results are shown in Table~2 for the 1997 observation
and in Table~3 
for the 1999 observation. In Fig. 4 we show the blackbody radius
$R_{\rm {BB}}$, the temperature $kT_{\rm {BB}}$, luminosity
$L_{\rm {BB}}$ and the power law photon index $\Gamma $ as a function of
$L_{\rm {Tot}}$, for the 1997 and 1999 observations, respectively.
%As is usual, errors are not quoted or
%plotted for $L_{\rm {CPL}}$ or $L_{\rm {Tot}}$, although these can be estimated  
%from the 90\% uncertainties in the CPL normalization of $\sim $15\%,
%plus systematic errors due to uncertainty in the high energy spectrum.

%%\tabcolsep 2.
\begin{table} %Table 2
\caption{Spectral fitting results for the nine 1997 {\it RXTE} PCA spectra
fitted simultaneously; $\chi^2$/dof = 400/405. The first row shows
the quiescent source, and subsequent rows show the development of flaring.
90\% confidence errors are given: luminosities are in units of 10$^{37}$ erg s$^{-1}$}
\begin{center}
\begin{tabular}{crccr}
\hline\noalign{\smallskip}
$kT_{\rm {BB}}$ & $R_{\rm {BB}} \; \; \;$  &$L_{\rm {BB}}$
& $\Gamma $&$L_{\rm {CPL}}$\\
keV & km $\; \; \;$&  erg s$^{-1}$&  &  erg s$^{-1}$\\
\noalign{\smallskip\hrule\smallskip}
$1.34\pm 0.02$&$10.2\pm 0.4$&$4.3\pm 0.2$&$2.09\pm 0.06$&7.6\\
$1.40\pm 0.03$&$8.9\pm 0.4$&$4.0\pm 0.2$&$2.04\pm 0.06$&8.1\\
$1.45\pm 0.03$&$8.3\pm 0.4$&$4.0\pm 0.2$&$1.98\pm 0.06$&8.3\\
$1.54\pm 0.05$&$7.3\pm 0.5$&$3.8\pm 0.3$&$1.90\pm 0.07$&8.8\\
$1.59\pm 0.06$&$6.7\pm 0.6$&$3.8\pm 0.3$&$1.88\pm 0.08$&9.6\\
$1.78\pm 0.08$&$5.4\pm 0.5$&$3.8\pm 0.3$&$1.91\pm 0.10$&9.8\\
$1.80\pm 0.07$&$5.4\pm 0.5$&$3.8\pm 0.3$&$1.85\pm 0.08$&10.0\\
$1.90\pm 0.07$&$4.8\pm 0.4$&$3.9\pm 0.3$&$1.78\pm 0.07$&10.9\\
$1.98\pm 0.10$&$4.5\pm 0.5$&$4.1\pm 0.4$&$1.79\pm 0.10$&11.1\\
\noalign{\smallskip}\hline
\end{tabular}
\end{center}
\label{tab:spec_paras}
\end{table}
%%\tabcolsep 2.0
\begin{table}[!hb]
\caption[ ]{Equivalent results for the 1999 data.
$\rm {\chi^2/dof}$ = 354/546 for 12 spectra fitted
simultaneously; first row is non-flare as before}
\begin{center}
\begin{tabular}{crccr}
\hline\noalign{\smallskip}
$kT_{\rm {BB}}$ & $R_{\rm {BB}} \; \; \;$  &$L_{\rm {BB}}$
& $\Gamma $ &$L_{\rm {CPL}}$\\
keV & km $\; \; \;$&  erg s$^{-1}$&  &  erg s$^{-1}$\\
\noalign{\smallskip\hrule\smallskip}
$1.37\pm 0.04$& $7.1\pm 0.5$ & $2.3\pm 0.2$ & $2.20\pm 0.04$&11.2\\
$1.37\pm 0.03$& $7.1\pm 0.4$  & $2.3\pm 0.2$ & $2.17\pm 0.03$&11.4\\
$1.42\pm 0.03$& $6.7\pm 0.4$  & $2.4\pm 0.2$ & $2.13\pm 0.03$&11.5\\
$1.50\pm 0.03$& $5.9\pm 0.3$  & $2.3\pm 0.1$ & $2.12\pm 0.03$&11.9\\
$1.62\pm 0.04$& $5.1\pm 0.3$  & $2.3\pm 0.1$ & $2.09\pm 0.03$&12.1\\
$1.71\pm 0.05$& $4.5\pm 0.3$  & $2.2\pm 0.1$ & $2.05\pm 0.03$&12.3\\
$1.89\pm 0.06$& $3.8\pm 0.3$  & $2.3\pm 0.1$ & $2.04\pm 0.04$&12.6\\
$2.02\pm 0.06$& $3.4\pm 0.2$  & $2.6\pm 0.2$ & $2.01\pm 0.04$&12.5\\
$2.16\pm 0.06$& $3.1\pm 0.2$  & $2.8\pm 0.2$ & $2.01\pm 0.15$&12.7\\
$2.18\pm 0.06$& $3.1\pm 0.2$  & $2.8\pm 0.2$ & $1.95\pm 0.05$&12.8\\
$2.22\pm 0.07$& $3.1\pm 0.2$  & $3.0\pm 0.3$ & $1.94\pm 0.06$&12.8\\
$2.25\pm 0.10$& $3.2\pm 0.4$  & $3.4\pm 0.3$ & $1.96\pm 0.11$&13.0\\
\noalign{\smallskip}\hline
\end{tabular}
\end{center}
\end{table}

In both observations, it was clear that strong changes take place during flaring in 
the blackbody component with $kT_{\rm {BB}}$ increasing from 
$\sim $ 1.3 keV to more than 2 keV. Remarkably, the blackbody radius
(where the emitting area is $4\,\pi\,R_{\rm {BB}}^{\rm {2}}$) {\it decreased} 
strongly in both observations. In the 1997 data, $L_{\rm {BB}}$ remained approximately
constant during flaring, and similar behaviour was seen in the 1999
data for $L_{\rm {Tot}}$ $<$ $\rm {1.5\times 10^{38}}$ erg $\rm
{s^{-1}}$ (equal to the largest $L_{\rm {Tot}}$ in 1997); at higher
$L_{\rm {Tot}}$, $L_{\rm {BB}}$ increased to $\rm {\sim 3.5\times 10^{37}}$
erg $\rm {s^{-1}}$ at the peak of flaring. Thus, it is clear that over
much of flare development, the decrease of $R_{\rm {BB}}$ offsets the
increase of $kT_{\rm {BB}}$ so that $L_{\rm {BB}}$ remains constant.
A consequence of the constancy of $L_{\rm {BB}}$ is that the increase in 
total luminosity is entirely in the Comptonized emission.
The power law photon index $\Gamma $ decreased systematically by 15\%
during flaring as seen in Fig. 4. 

The iron line had a non-flaring EW of 60$\pm$ 20 eV in the 1997 data, 
and 90$\pm$ 30 eV in the 1999 data. We do not show 
\begin{figure}[!ht]
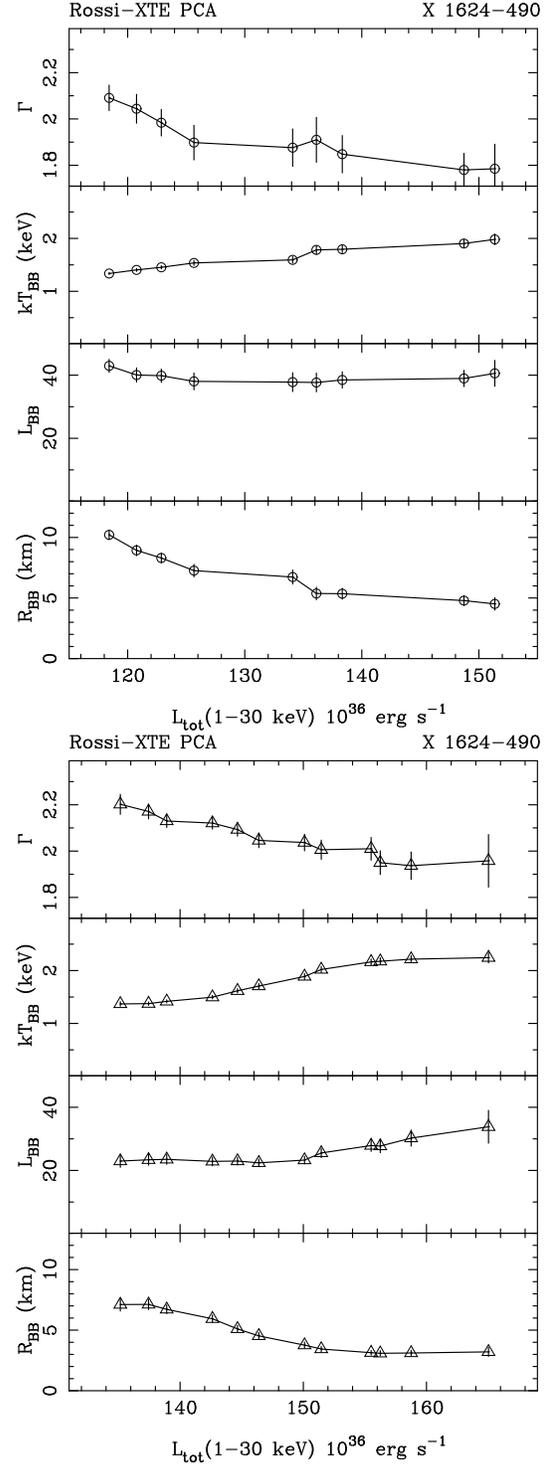
 %Fig. 4
\begin{center}
\includegraphics[width=70mm,angle=0]{f4a} % orig spec_res_jan97 76
\includegraphics[width=70mm,angle=0]{f4b} % orig spec_res_sep99
\caption{Upper: Spectral fitting results from the 1997 observation:
blackbody parameters and power law photon index $\Gamma $. The blackbody
luminosity is in units of ${\rm 10^{36}}$ erg s$^{\rm -1}$.
Lower: Corresponding results from the 1999 observation}
\end{center}
\end{figure}
EW in Tables 2 and 3 for the growth of flares,
because the continuum flux is changing; the line intensity is more relevant.
\begin{figure}[!h]
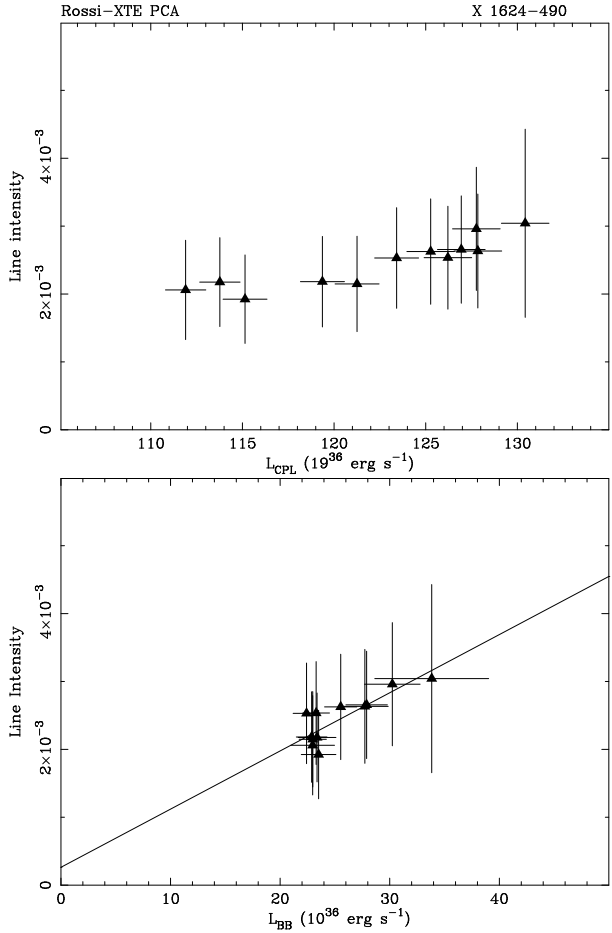
 %Fig. 5
\includegraphics[height=80mm,angle=270]{f5a} % orig pl.qdp
\includegraphics[height=80mm,angle=270]{f5b} % orig line.qdp
\caption{Variation of iron line intensity (in photon cm$^{-2}$ s$^{-1}$)
with the luminosity of the Comptonized
emission (upper panel), and with the blackbody luminosity (lower panel).
The best-fit line is shown in the lower panel (see text)}
\end{figure}
In the 1997 data, the intensity of the 6.4 keV line did not change, and it will
be relevant that $L_{\rm BB}$ was also constant. In the 1999 data, the line intensity
$I_{\rm line}$ did change and is plotted against $L_{\rm BB}$ and $L_{\rm CPL}$ (the luminosity of
the Comptonized emission) in Fig. 5. This shows that during the initial growth of flaring the
line intensity does not change significantly although $L_{\rm CPL}$ is increasing; $L_{\rm
BB}$ is constant at this stage. However, in the second stage of flare growth,
$L_{\rm BB}$ increases (Fig. 4), and then the line intensity increases. In Fig. 5, the
best-fit line to the variation of $I_{\rm line}$ with $L_{\rm BB}$ is
shown, this taking into account the 90\% confidence errors in both $y$ and $x$.
This line has gradient ${\rm 8.57\pm 0.69\times 10^{-5}}$ photon
cm$^{-2}$ erg$^{-1}$, showing that
the line intensity is correlated with the blackbody luminosity with
high significance. This is the first time that a line variation has been linked to the variation
of a particular continuum emission component; this will be discussed further
in Sect. 4.

The data of Fig. 4 are replotted in Fig. 6 as $kT_{\rm BB}$ against $L_{\rm Tot}$ for both
{\it RXTE} observations, plus a point 
from the {\it BeppoSAX} observation 
(Ba\l uci\'nska-Church et al. 2000), when the source was substantially less bright. 
We have made comparisons of {\it RXTE } PCA Crab data in the band 3--20 keV 
(obtained before and after the 1999 observation of X\th 1624-490) with {\it BeppoSAX} 
MECS data of 1998, October, which shows that systematic differences between flux from the 
two instruments amount to no more than 17\%, (higher in {\it RXTE}).
Fig. 6 shows that at the peak of flaring in the 1999 data, $kT_{\rm
{BB}}$ saturates at a constant value at the same time that $R_{\rm
{BB}}$ becomes constant at a low value (Fig. 4). During this stage of
flaring, $L_{\rm {BB}}$ increases, eventually becoming fixed
when the emitting area and $kT_{\rm {BB}}$ become constant.
In Fig. 7, $L_{\rm {BB}}$ is plotted against blackbody temperature
$kT_{\rm {BB}}$. The approximate constancy of $L_{\rm {BB}}$ 
in the early stages of flaring is apparent. Also shown
is a line drawn through the penultimate point of the 1999 observation
before the peak of flaring, representing the relation $L_{\rm {BB}}$ =
$A\, \sigma \, T^{\rm {4}}$. In this second stage of flaring as
$R_{\rm {BB}}$ becomes constant, further increase in $L_{\rm {BB}}$  
is due to temperature increase at $\sim $ constant area, so that the
data obey Stefan's Law.

\begin{figure}[!h] %Fig. 6
\includegraphics[width=66mm,angle=270]{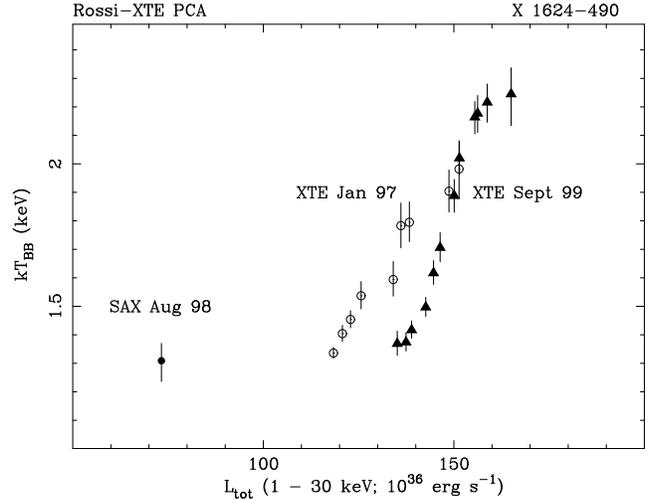} % orig Ltot_temp
\caption{Variation of blackbody temperature $kT_{\rm {BB}}$ with
source luminosity for the 1997 and 1999 {\it RXTE} observations and the
1998 {\it BeppoSAX} observation}
\end{figure}

\begin{figure}[!h] %Fig.7
\includegraphics[width=66mm,angle=270]{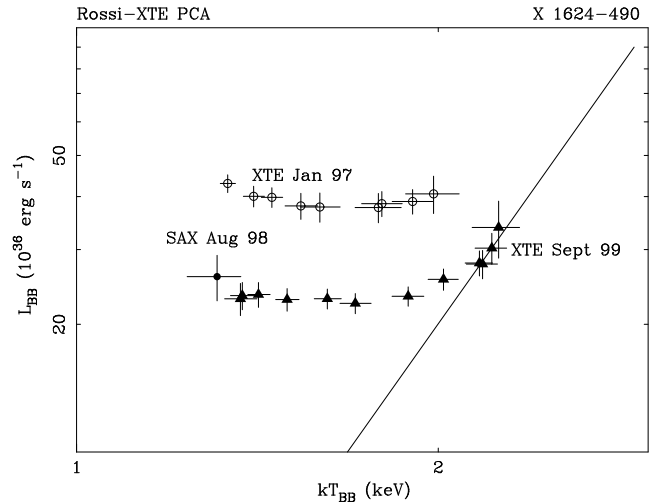} % orig Lbb_temp
\caption{Blackbody luminosity as a function of blackbody temperature
which increases as flares evolve. The constancy of $L_{\rm {BB}}$
over most of the flaring can be seen, and the final increase of
luminosity agreeing with a $T^4$ law, as the blackbody emitting area
becomes
constant (see text)}
\end{figure}

\begin{figure}[!h]           %Fig. 8
\includegraphics[width=66mm,angle=270]{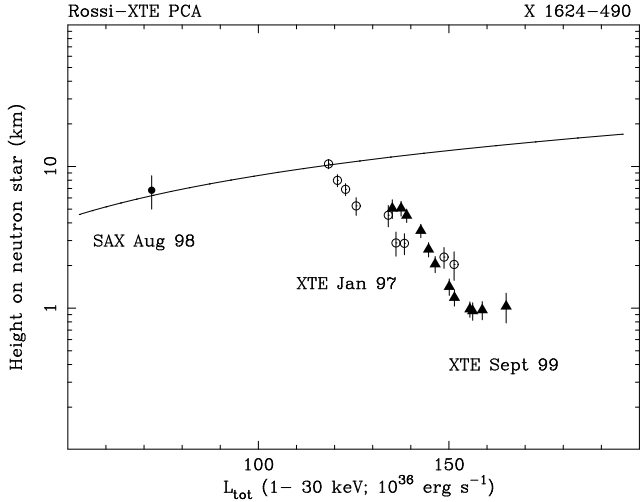}     %orig hh_final
\vskip 10mm
\caption{Evolution of the blackbody emitter during flaring in X\th
1624-490; the curve shows the equilibrium height of the
radiatively-supported inner disk $H_{\rm {eq}}$}
\end{figure}
 
Next, the results are re-expressed in the way used by Church \&
Ba\l uci\'nska-Church (2001) in a survey of LMXB sources using {\it ASCA} and 
{\it BeppoSAX}. In Fig. 8 we show the variation of the half-height
{\it h} of the emission region on the neutron star as a function of
$L_{\rm {Tot}}$. It is assumed that emission is from an equatorial
strip on the star with area $4\,\pi\,h\,R$ (a sphere
intersected by two parallel planes) equal to 
$4\,\pi\,R_{\rm {BB}}^{\rm {2}}$. Thus, {\it h} = $R_{\rm {BB}}^{\rm 2}/R$, 
and it is assumed that $R$ is 10 km. 
Also shown are calculated values of the equilibrium half-height $H_{\rm
{eq}}$ of the inner radiatively-supported accretion disk given
approximately by the equation (Frank, King \& Raine, 1992):

\begin{equation}
H = {3\,\sigma\,\dot M\over 8\,\pi m_{\rm {p}}\,c} \left\lbrack 1 - \left({R\over 
r}\right)^{\rm {1/2}}\right\rbrack = H_{\rm {eq}} \left\lbrack 1 -
\left({R\over r}\right)^{\rm {1/2}}\right\rbrack
\end{equation}

\noindent
This equation shows that the disk height increases rapidly with radial
distance from zero at the surface of the neutron star towards the equilibrium 
value $H_{\rm {eq}}$. Most of the increase takes place between r = 10 and r = 20 km, 
i.e. within 10 km of the stellar surface.
In weak sources, with $L$ $\sim 5\times 10^{\rm {36}}$
erg $\rm {s^{-1}}$, radiative support is over a limited radial extent
so that $H_{\rm {eq}}$ is never achieved. However, in bright sources
with $L$ $\rm {\sim 10^{38}}$ erg $\rm {s^{-1}}$, such as
X\th 1624-490, the equilibrium height is achieved, and the
radiatively-supported disk extends to a radial distance of $\sim $400 km.
For the sources in the LMXB survey there was found to be an agreement {\it h = H}
spanning 3 orders of magnitude in luminosity.

In the case of X\th 1624-490, we
calculated $H_{\rm {eq}}$ from $L_{\rm {Tot}}$ using the expression
$L_{\rm {Tot}}$ = $G\, M\,\dot M/ R$, where $M$ is the mass of
the neutron star taken to be 1.4$\rm {M_{\sun}}$.
However the increase of luminosity in flaring probably does not
mean an increase in $\dot M$ (as in X-ray bursts). Thus the curve
of $H_{\rm eq}$ shown in Fig. 8 is appropriate for comparison of the
non-flaring source in the 3 observations, but should not be taken to
imply $\dot M$ increases in flares. 
It can be seen that there is good agreement between
{\it h} and $H_{\rm {eq}}$ for the {\it BeppoSAX} and 1997
{\it RXTE} observations. In the 1997 data, $h$ = 10 km, i.e. the
whole neutron star is emitting.
During flaring, there is a strong depression
of $h$ below $H_{\rm {eq}}$ by a factor of $\sim$5 at the peak of
flaring. In the 1999 observation, it can be seen that the non-flaring
source has $h$ already depressed below $H_{\rm {eq}}$.

As flaring develops in the 1997
observation, $L_{\rm {Tot}}$ increases from $\rm {1.15\times 10^{38}}$ erg
$\rm {s^{-1}}$ to $\rm {1.5\times 10^{38}}$ erg $\rm {s^{-1}}$. The
1999 {\it RXTE} data shows similar behaviour, except that the
non-flaring luminosity was higher at $\rm {1.3\times 10^{38}}$ erg 
$\rm {s^{-1}}$ rising to $\rm {1.7\times 10^{38}}$ erg s$^{-1}$ in flares. 
It is important to decide whether the initial depression of $h$ below
$H_{\rm {eq}}$ is real so the effects of uncertainty
in the Comptonization cut-off energy discussed previously were tested.
The value of $E_{\rm {CO}}$ was allowed to vary between its smallest
value 9 keV obtained in free fitting and the maximum value 21 keV.
The effect of this was that the total luminosity (1--30 keV) varied as might be expected,
from the best-fit value of 
$\rm {1.3\times 10^{38}}$ erg $\rm {s^{-1}}$ for $\rm {E_{CO}}$ = 11.8 keV by 
$\rm {\pm 0.1\times 10^{38}}$ erg $\rm {s^{-1}}$ for the
non-flaring spectrum. However, there was essentially no change in the
height of the emitting region on the neutron star. 
%Similar tests were %carried out for the 1997 data, with similar possible ranges for
%$L_{\rm {Tot}}$. %The best-fit value of {\it h} was 10 km, which could be increased
%by 20\% for $E_{\rm {CO}}$ = 5 keV (free fitting), but was not changed
%by increasing the value to 12 keV (the maximum allowed by fitting).
We can conclude that the dramatic changes shown in Fig. 7 are not
affected by this uncertainty.

\section{Discussion}

The main result of our analysis is a neutron star blackbody radius 
which decreases strongly as flares develop. We firstly consider whether
this can be influenced by the fact that we fitted a simple blackbody
when it is often assumed that X-ray burst spectra, particularly at the
peaks of bursts can have a modified blackbody spectral form, due to the
effects of electron scattering in the neutron star atmosphere. There
were two types of evidence for a modification taking place: blackbody
temperatures of 2.5--3.0 keV implying super-Eddington luminosities
in bursts, and an anti-correlation of blackbody radius $R_{\rm BB}$
with $kT_{\rm BB}$ in a particular source (below) when a simple
blackbody was fitted. It should be mentioned at this stage
that it would be unwise {\it not} to fit a simple blackbody model
to the present data as the spectral form depends on one parameter only
$kT_{\rm BB}$, allowing straightforward interpretation of the results.
Particular modified blackbody models have a flux depending on $kT$ and
electron density and so provide {\it no} information on the 
emitting area unless some assumption is made about the electron density.

In the study of X-ray bursts, it was first shown by Swank et al. (1977)
that the burst spectra were well described by a simple blackbody. This was supported
by the fact that during burst decay, the flux varied approximately as $T^{\rm 4}$
(Hoffman et al. 1977). By fitting a simple blackbody model, burst peak blackbody 
temperatures of $\sim$2 keV are generally found; however higher temperatures of 
$\sim$3 keV have been measured, also obtained using high quality 
data from the {\it Rossi-XTE} PCA (Swank priv. comm.), so that high temperatures at the peak 
of some bursts cannot be doubted. As these high temperatures
imply super-Eddington luminosities, theory was developed in which 
electron scattering dominated over free-free absorption in the opacity 
of the neutron star atmosphere during bursts, and the
surface value of the radiation source function departed from Planck
form and had a modified blackbody form. Modified blackbody spectral
forms have been given for a constant density atmosphere which we will
designate {\sc mbb1} (van Paradijs
1984) and for an exponential atmosphere {\sc mbb2} (Zel'dovich \& Shakura 1969).
Modelling of the neutron star atmosphere has been carried out by 
Fujimoto et al. (1981) who derived conditions for stable and unstable
hydrogen and helium burning, by Paczy\'nski (1983), by London et al. (1984,
1986), Ebisuzaki et al. 1984, and Ebisuzaki 1987. London et al. modelled
the atmosphere numerically allowing for the effects of Comptonization.
Although the simpler models {\sc
mbb1} and {\sc mbb2} have a spectral form which differs from the Planck
form, and could, in principle, be detected if sufficient count
statistics were available, the more sophisticated numerical models 
have forms which are similar to the Planck form although the peak energy
has a different relation to $kT$. However, when close to the Eddington
limit, a substantial low-energy hump is expected (Madej 1991).
In our work on LMXB,
we have never found any indication of a departure from simple blackbody
spectral shape in the quiescent spectra, all sources studied being well-fitted by
using a simple blackbody plus cut-off power law model (e.g. Church \& Ba\l uci\'nska-Church 
1995; Ba\l uci\'nska-Church et al. 1999; Smale et al. 2001).
X-ray burst spectra have been well-fitted by a simple blackbody, e.g. 
using the {\it Rossi-XTE} PCA with its high sensitivity, it has been possible to 
examine burst spectra accumulated over times
as short as 0.1~s, and these are well-fitted by a simple blackbody
as in 4U\th 1728-34 (Swank 2001). One possibility
remaining is that bursts with $kT$ $\sim$3 keV may have modified spectra, 
whereas those with $kT$ $\sim$2 keV do not.

\begin{figure}[!h]
\includegraphics[width=63 mm,angle=270]{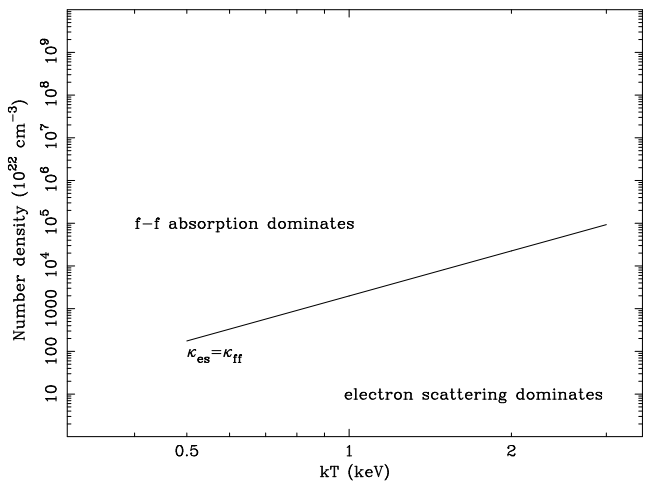}    % was dens.qdp
\caption{Relative contributions of absorption and electron scattering 
to the opacity in the neutron star
atmosphere as a function of number density and temperature $kT$}
\end{figure}

Modification of the blackbody spectrum was expected on the grounds
that in an X-ray burst, the density was ``low'' and the temperature
``high'', and
we next address the question of how well the electron density is
known. It is convenient to use a 
density-temperature diagram on which we can delineate the regions in which free-free and 
electron scattering processes dominate as shown in Fig. 9.
Density can be expressed as electron number density, as mass density $\rho $, 
or as the vertically integrated mass column density $\Sigma $.
The problem is that the value of density used in various theoretical treatments 
of X-ray bursting are very different. Fujimoto et al. (1981) provide values of 
$\Delta$M $\sim \rm {10^{-12}}$--${\rm 10^{-11}}$ M$_{\sun}$ for the total mass needed to be 
accumulated on the surface of the star for a nuclear flash to occur. From this, 
$\Sigma$ = 10$^8$--10$^9$ g cm$^{-2}$ and if we assume the thickness of the atmosphere 
in the range 10--100 cm, the average $\rho$ is 10$^6$--10$^8$ g cm$^{-3}$. However, 
the density can vary considerably from the mean, and be different in the hydrogen 
and helium burning shells (Joss \& Rappaport 1984). London et al. (1984) assume a 
value of $\Sigma $ = 100 g cm$^{-2}$. Ebisuzaki (1984) gives a range of mass densities 
in the surface layer for a burst luminosity of 20\% of Eddington from 1--$\rm {10^4}$
g cm$^{-3}$, which correspond to $\Sigma $ in the range 5--$\rm {5\times 10^4}$ g
cm$^{-2}$. Paczy\'nski (1983) used values of surface mass ($\Delta$M) similar to 
those of Fujimoto et al. (1981). Thus these values correspond to $\Sigma$ spanning 
the range 5--10$^9$ g cm$^{-2}$ and to electron number densities $n_{\rm e}$ in 
the range 10$^{24}$--10$^{32}$ cm$^{-3}$. While an atmosphere with $n_{\rm e}$ =
10$^{24}$ cm$^{-3}$ is clearly dominated by electron scattering, this is not the 
case for higher values of electron density. The calculation of densities in neutron 
star atmospheres during bursts is complex and it does not appear possible yet to be 
definite about values. 

The second piece of evidence for modification of the blackbody 
in X-ray bursts was the apparent decrease in blackbody radius with increasing 
temperature $kT$ when a simple blackbody model was fitted to {\it Tenma} and 
{\it Exosat} data of bursts in XB\th 1636-536 (Inoue et al. 1984; Sztajno et al. 1985). 
In these data, $R_{\rm BB}$ decreased 
\begin{figure}[!ht]            %Fig. 10
\begin{center}
\includegraphics[width=63mm,angle=270]{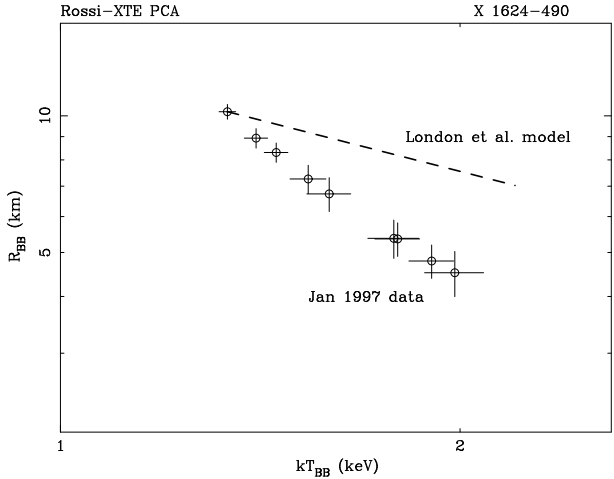}             % was rt.qdp
\caption{Test of the hypothesis that the apparent decrease in blackbody 
area in flaring is not real by comparing the 1997 results with the 
$R$-$T$ relation predicted by the London et al. model atmosphere (dashed line) 
assuming that the emitting area remains constant (see text)}
\end{center}
\end{figure}
by about 40\% as $kT_{\rm BB}$ increased from 1 to 2 keV. Sztajno et al. 
showed that the apparent decrease was consistent with a constant
emitting area if the model atmospheres of London et al. (1984, 1986)
were used. We will make the same comparison with the present data.
London et al. construct models for various values of the emitted flux $f$, which
they re-express in terms of an effective temperature $T_{\rm eff}$,
defined {\it via} $f$ $\sim$ $T_{\rm eff}^4$. Fitting a blackbody to the
emission of their model atmospheres provides a spectral or colour
temperature $T_{\rm spec}$, so that if the luminosities of the fitted
blackbody and generated spectrum are equal: 
\begin{equation}
4\, \pi\, R_{\rm BB}^2\, \sigma\, T_{\rm spec}^{\rm 4} = const \cdot R^2 \, T_{\rm eff}^{\rm 4}
\end{equation}

\noindent Assuming $R$ is constant gives
$R_{\rm BB}$ $\propto$ $(T_{\rm eff}/T_{\rm spec})^2$. 
Using the spectral hardening factor $T_{\rm spec}/T_{\rm eff}$ 
$\sim$ $T_{\rm eff}^{\rm 3/5}$ of London et al., we
obtain $R_{\rm BB}$ $\sim $ $1/T_{\rm spec}^{\rm 3/4}$
for the expected dependence of the apparent decrease in blackbody
radius with increasing measured burst temperature. In the case of the present
flare data if we assume some modification of the spectrum takes place,
as $kT$ increases from 1.34 to 1.98 keV (1997 data), we would expect a decrease
in $R_{\rm BB}$ of 34\%, whereas the actual increase is by a factor of
2.3, and these changes are compared in Fig. 10.
Thus, it does not seem very likely on the basis of the London
et al. model that the emitting area can actually
remain constant while appearing to decrease by this large factor.

If we use the simpler {\sc mbb1} modified blackbody model,
the luminosity is related to the electron density {\it via}
\hbox{$const\cdot R^{\rm 2}\,(kT)^{\rm 9/4}\, n_{\rm e}^{\rm 1/2}$}
(with a similar relation for {\sc mbb2}).
To demonstrate the constraints imposed on $n_{\rm e}$, we fitted the 1999
data using the model {\sc mbb1 + cpl}. This fitting provided good
values of $\chi^{\rm 2}$/dof; results are shown in Table 4.
We assume that the spectrum of the non-flaring source
is an unmodified blackbody, but that mid-flare and peak-flare can be
modified. In the non-flare emission, the blackbody radius $R_{\rm BB}$ 
was 7.1 km. Assuming that this does not change in flaring, we use the measured
{\sc mbb1} normalization and $kT$ values to derive the electron density at the peak of
flaring, $n_{\rm e}$ = ${\rm 1.9\times 10^{23}}$ cm$^{-3}$.
Not only is this fine-tuning of electron density values, but the value is relatively
low in terms of expected values in model atmospheres as discussed above.
Moreover, the electron density would have to decrease by a factor of two 
between mid-flaring and the peak of flaring to keep $R$ constant. 
It is unlikely that $n_{\rm e}$ decreases in flaring and so 
for this simple model also, it is unlikely that the emitting
area does not decrease.

%%\tabcolsep 3.0
\begin{table}[!hb]
\begin{center}
\caption{Results of fitting non-flare emission by simple blackbody and
flaring by a modified blackbody spectral form. {\it norm} has units of $\rm {10^{37}}$ erg s$^{-1}$ 
at 10 kpc; electron densities are in units of 10$^{23}$ cm$^{-3}$}
\begin{flushleft}
\begin{tabular}{llrr}
\hline\noalign{\smallskip}
$\;\;\;$&Non-flare & Mid-flare  &Peak-flare \\
\noalign{\smallskip\hrule\smallskip}
& Simple BB & Modified BB & Modified BB\\
$kT$ (keV)    & 1.37$\pm$0.04 & 2.30$\pm$0.07  &3.46$\pm$0.10\\
norm  &$\rm {1.02\pm 0.09}$ &$\rm {1.86\pm 0.11}$&$\rm {3.35\pm 0.53}$\\
$R$ (km)      &7.1$\pm$0.5 & 7.1 & 7.1\\
$n_{\rm e}$ &\dots &$\rm 3.7\pm 0.7$&$\rm 1.9\pm 0.3$\\
\noalign{\smallskip}
\hline
\end{tabular}
\end{flushleft}
\end{center}
\end{table}

However, it is not proven that a modification does take place in
flaring, and there are arguments that it does not. Firstly, 
the value of $kT$ $\sim $2 keV obtained at the peaks of flares does not demand a 
modification of the spectrum in the way that super-Eddington values of $kT$ = 3 keV 
do in X-ray bursts. Secondly, X-ray flaring is much less well understood than X-ray 
bursting and so our knowledge of densities in the neutron star atmosphere is even 
more uncertain as no theoretical model of the atmosphere during flaring exists at 
present. We have previously suggested (Sect. 3.1) that an X-ray flare consists of a 
superposition of X-ray bursts. The recurrence time between these is expected to be 
short as observed, because of the much higher values of mass accretion rate compared 
with a burst source. The 1--30 keV luminosity of \src\ in the observations discussed 
here is $\sim 10^{38}$ erg s$^{-1}$, i.e. considerably higher than that of a typical 
burst source. Consequently it is likely that the densities in the atmosphere are more 
than 10 times higher making it less likely that electron scattering will dominate the 
opacity. In summary, there is no observational evidence of a
modification of the blackbody spectrum. However, if a 
modification did take place, it is unlikely that this can explain the
large observed decrease in area, and we conclude that the main results of 
the analysis made using a simple blackbody are correct.

\begin{figure}[!h] %Fig. 11
\includegraphics[width=65mm]{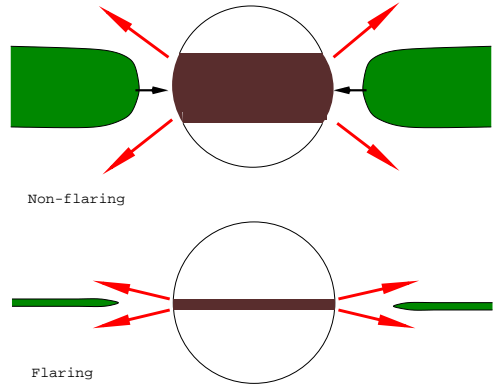}      % = was [combined]
\caption{Schematic diagram showing the result from spectral fitting that the emitting area 
of the neutron star blackbody decreases strongly during flaring. The top panel shows the 
calculated height {\it H} of the radiatively-supported inner disk which agrees well with 
the neutron star emitter height {\it h} in the non-flaring state. One possibility (see text) 
is that the vertical extent of the inner disk is decreased in flaring
(lower panel)}
\end{figure}

We now discuss possible reasons for the decrease in emitting area.
Results of the {\it ASCA} survey of LMXB (Church \& Ba\l uci\'nska-Church 2001)
showed that the level of neutron star blackbody emission can be
explained by assuming that the half-height of emitter on the star $h$
is equal to the half-height $H$ of the inner radiatively-supported disk
(see Sect. 1). Two possible explanations exist (see Sect. 1):
either there is radial flow across the gap between disk and star (mapping
the value of $H$ onto $h$), or $H$ is simply a measure of $\dot M$,
and the mechanism of Inogamov \& Sunyaev (1999)
(accretion flow spreading on the stellar surface) determines $h$.
If the first of these is correct, then radiation
pressure can affect the inner disk by material being removed at the upper
and lower surfaces by radiation pressure, leading directly to a decrease in {\it h}
and of $\dot M$ reaching the star 
in flaring, as shown in Fig. 11. If the second mechanism is correct,
then we need the accretion flow between disk and star to be impeded by radiation pressure.
In either case, it is important to compare the levels of blackbody 
emission obtained from spectral fitting with the Eddington limit to test whether the 
effects of radiation pressure should be expected. If the emitting area is reduced at the 
peak of flaring to a narrow equatorial strip, radiation will not be isotropic. For an 
observer at angle $\theta$ from the equatorial plane, the observed flux will be $f_{\rm obs}$ 
= $f\cdot {\rm cos}\theta $ where $f$ is the flux in the plane. For X\th 1624-490, $\theta$ 
$\sim$30$\degmark$. To test whether Eddington limiting takes place in the equatorial 
plane, we should compare $f$ = $f_{\rm obs}/{\rm cos}\theta$ with $f_{\rm Edd}$.

Taking into account the
non-Newtonian gravitation near the neutron star and possible variations in the composition of
neutron star atmosphere so that the opacity may not equal the Thompson scattering opacity,
there will be a local value of critical luminosity
$L_{\rm crit}$ = \hbox{$L_{\rm Edd}\cdot {\kappa \over\kappa_{\rm {T}}}\cdot (1+z)$}
where $L_{\rm Edd}$ is the normal Eddington limiting luminosity, 
$\kappa _{\rm T}$ is the Thomson scattering opacity, $\kappa $ is the actual opacity
and $z$ the gravitational redshift
(Paczy\'nski 1983). For a distant observer, this critical value
will be reduced by the gravitational redshift to $L_{\rm crit,\infty}$, so that

\begin{equation}
L_{\rm crit,\infty } = {4\pi\, G\, M\,c\,\over \kappa_{\rm
T}}\cdot{\kappa_{\rm T} \over \kappa}\cdot {1\over (1 + z)}
\end{equation}

\noindent
where 
$4\pi\, G\, M\,c\,/ \kappa_{\rm T}$ is the standard Eddington
limiting luminosity $L_{\rm Edd}$ = $\rm {1.76\times 10^{38}}$ erg s$^{-1}$
for a 1.4 M$_{\sun}$ neutron star. This depends on composition since
$\kappa_{\rm T}$ $\simeq {\rm 0.2 (1 + X)}$ where X is the fraction by
weight of hydrogen. The factor
$1/(1+z)$ = $\sqrt{1 - 2GM/Rc^{\rm 2}}$ in terms of the mass $M$ and radius $R$ of 
the star. We will compare our measured blackbody luminosities with the 
Eddington limit by comparing with $L_{\rm crit,\infty}$, assuming simply
hydrogen composition and a value of $z$ = 0.34 appropriate to the 
surface at $R$ = 10 km of a 1.4 M$_{\sun}$ neutron star. Thus
$L_{\rm crit,\infty}$ = $\rm {1.31\times 10^{38}}$ erg s$^{-1}$. 

In both the 1997 and 1999 data, the observed blackbody luminosity $L_{\rm BB}$ remains 
less than $\rm {4\times 10^{37}}$ erg s$^{-1}$. Thus the flux in the
equatorial plane corresponds to a luminosity 
of $\rm {\sim 5\times 10^{37}}$ erg s$^{-1}$, i.e.
about a factor of 2.5 less than $L_{\rm crit,\infty}$. This
calculation suggests that the effects of radiation pressure are
important. In the 1997 data, $L_{\rm BB}$ remains about constant at 
$\rm {4\times 10^{37}}$ erg s$^{-1}$ and the blackbody radius $R_{\rm BB}$
decreases from its non-flaring value of $\rm {10.2\pm 0.4}$ km
as $kT_{\rm {BB}}$ increases. In the 1999 data, 
$R_{\rm {BB}}$ starts at a somewhat lower level than in 1997 at 7.1 km 
so that the effects seen in flaring are already taking place in the non-flaring emission.
The effects of radiation pressure clearly extend over a range of blackbody 
luminosity values, but at the peak of flaring further increase of 
$L_{\rm {BB}}$ is prevented due to both $kT_{\rm {BB}}$ and $R_{\rm {BB}}$ 
reaching stable values. Exactly how the lower limit of $R_{\rm {BB}}$ is caused is not
clear. The processes discussed of increasing $kT_{\rm BB}$ and decreasing $R_{\rm BB}$ 
constitute a mechanism for limiting the blackbody luminosity and thus 
an Eddington limiting effect not previously known. At the peak of flaring, the 
mass accretion flow to the star will be reduced below the non-flaring case by 
the changes in the inner disk and star, the radiation pressure will fall, and 
the height of the emitting region on the star increase.
Thus, the flare is self-limiting and the system will return to the
non-flare state on a similar timescale to that of flare growth.

We next discuss further the implications of the relation {\it h = H} obtained from the {\it
ASCA} and {\it BeppoSAX} survey (Church \& Ba\l uci\'nska-Church 2001),
and supported by the present work and the two possible explanations
referred to previously. The first explanation involves radial flow between
inner disk and star. Although there has been a large
amount of theoretical work on advective flow in black hole binaries,
e.g. Abramowicz et al. (1996), and references therein, and on the
nature of the disk itself, such as on slim disks (Abramowicz et al.
1988; Igumenschev et al. 1998), there has been relatively little corresponding
work on the accretion flow between the inner disk and neutron star 
in LMXB. It is not known whether the sonic point lies
within the star or within the inner disk. However, Popham \& Sunyaev
(2001) investigate the nature of the inner accretion disk around a
neutron star, assuming a boundary layer within the disk, and show that the radial velocity 
increases by about two orders of magnitude leading to advective flow within 
the disk and increase in disk height due to increased heating. This
does not answer the question, of course, as to whether material can
flow across the gap between the inner disk edge and the star, although
Popham \& Sunyaev comment that they would intuitively expect material to fall onto the
star in a wide belt. Two- or three-dimensional modelling may reveal
whether such radial flows are possible. It has been suggested that
the model of radial flow between an inner radiative disk and the star 
is invalid as the inner disk may be unstable
to the radiation pressure instability discussed by Janiuk et al. (2000)
for a 10 M$_{\sun}$ black hole. However, scaling from a 
10 M$_{\sun}$ black hole to a 1.4 M$_{\sun}$ system indicates an
instability timescale of $\sim$30 s, and this, even if present, 
may not disrupt the disk.

The alternative explanation of our results
involves the flow of accretion material
on the surface of the neutron star as proposed by Inogamov \& Sunyaev
(1999). The ageement between 
$h$ and $H$ would result from {\it H} being a measure of the total
luminosity, not from $H$ {\it directly} determining $h$ in some way.
However, preliminary results of a comparison of this model with the survey results
show that the observed blackbody luminosities are $\sim$3 times larger 
than predicted by the theory (Church et al. 2001). This implies therefore, 
that radial flow may be the dominant factor determining the neutron star blackbody
emission. 
The present results on flaring suggest for the radial flow explanation that radiation
pressure can reduce the inner disk height so that both the emitting
area on the star, and the mass accretion rate reaching the star are
reduced. Possibly, the accretion flow could be affected without the
disk height changing. In the case of the Inogamov \& \hbox{Sunyaev}
mechanism, the decrease of $\dot M$ would lead to a decrease of
emitting area.

The present results are difficult to reconcile with the
possibility that the blackbody emission in LMXB could originate in the
accretion disk as multi-colour disk (MCD) blackbody emission. There
would be no reason to expect any agreement between $H$ and $h$, as $h$
would be related to the inner radius $r_{\rm i}$ of the disk blackbody
({\it via} $h$ = $r_{\rm i}^{\rm 2}/R$). The survey of LMXB (Church \& Ba\l
uci\'nska-Church 2001) showed that for the majority of sources, a Comptonization plus MCD
blackbody model required unphysical values of inner disk radius, substantially less
than the radius of the neutron star. The absence of disk blackbody
emission in LMXB can however, be understood in terms of the very
extended flat, nature of the ADC revealed by dip ingress time
measurements (Sect. 1). An extended ADC of radius typically 50,000 km
and large optical depth (Church 2001) will completely cover the hot X-ray
emitting inner accretion disk, so that all thermal emission will be
Comptonized and no disk blackbody component seen.

Finally, we will discuss the broad iron line at $\sim $6.4 keV. In Paper I, 
we reported the detection of this line in non-dipping, non-flaring emission. 
It has an equivalent width of $\sim$60 eV and $\sim $90 eV for the 1997 and 
1999 observations, respectively. Asai et al. (2000) made the first detection
of a broad line in this source at 6.4$^{+0.1}_{-0.3}$ keV using {\it ASCA} data;
with an EW of $13^{+13}_{-9}$ eV.
In the present paper, we show that the line intensity is
proportional to the luminosity of the blackbody component, the first
demonstration that a line intensity is linked to a particular component
of the continuum. Iron line emission is not well understood in LMXB
with conflicting models for line origin. Asai et al. (2000) from their 
extensive study of iron lines in LMXB with {\it ASCA} find that the line energy 
had a mean value of 6.57 keV in 20 sources and argue that lines are formed
by radiative recombination of photoionized plasma in the ADC, also proposed
by Hirano et al. (1987) and  White et al. (1985, 1986). In many sources, the line 
energy is $\sim $6.7 keV strongly suggesting a blend of Fe XXV lines at 6.63, 
6.67 and 6.70 keV formed by recombination. Although there is some scatter in the 
line energies between extremes of 6.2 and 7.0 keV it is thought likely that the 
lines in all sources have a common origin. In general, measurement of line
energies of 6.4--6.5 keV implying fluorescence has been taken to indicate
origin in the accretion disk (e.g. Barret et al. 2000), although the number
of detections of iron lines at these energies in LMXB is small.
In the case of X\th 1624-490, the variation of the line in dipping
(Paper I) is well described by giving the line the same covering factor as the ADC
which supports the origin of the line in the ADC although the measured energy
is unusually low at 6.4 keV. For the line energy to peak at 6.4 keV, 
an ionization parameter $\xi$ = $L/n\,r^{\rm 2}$ $\sim$100 is 
required which may be possible if the density in the ADC is high.
However, it is difficult to reconcile this line with the theoretical
expectation that the electron temperature in the ADC of several keV
might be expected to prevent formation of a 6.4 keV line.

An iron line produced in the ADC will have no bearing on the
non-detection of a reflection component in this source.
In fact, there has been a general lack of detections of reflection 
components in LMXB and reported detections in a small number of sources have been 
ambiguous. Barret et al. (2000) presented evidence for a weak reflection 
component in GS\th 1826-238 and SLX\th 1735-269. However, in the first case, 
the spectrum was modelled with a one-component Comptonization term only
but without a thermal component, which our previous work, e.g.
the {\it ASCA} survey
of LMXB (Church \& Ba\l uci\'nska-Church 2001) indicates
is always present and would be detectable at the measured
luminosity of the source. In the second case, the quality of fit
was acceptable ($\chi^2$/dof = 45/59) before a reflection
component was added. In a {\it Ginga} observation of XB\th 1608-522,
broad residuals above 7 keV could be modelled either by partial
absorption or a reflection component (Yoshida et al. 1993).
In all of our previous work using the {\sc bb + cpl} model,
there has been no trace of broad residuals indicating the presence of an
additional continuum component. We can suggest a possible reason for the lack of detections. 
We have demonstrated
the very extended, flat nature of the ADC (Church 2001; Smale et al. 2001),
with radius $r_{\rm ADC}$ typically 50,000 km, or 15\% of the disk radius,
and particularly large in \src\ (see Sect. 1). Thus
the ADC will cover all of the inner disk and 
act as a ``hot'' reflector preventing illumination
of the disk by the neutron star so preventing reflection.
In addition, the large values of optical depth obtained for the ADC
(Church 2001) will mean that any reflection component due to
illumination by the ADC itself will not penetrate the ADC to the
observer. In Cyg\th X-1, $r_{\rm ADC}$ is much smaller (Church 2001) and there is evidence for
a reflection component, suggesting that LMXB neutron star binaries
may differ from Black Hole binaries in this respect due to the strong effects of the neutron 
star in forming an extended ADC.

Future work will concentrate on further investigation of flaring in
sources such \src\ and on the flaring branch of Z-track sources. It is
clear that spectral evolution during flaring can provide valuable information
on the inner disk/neutron star interface not otherwise available.

%\begin{acknowledgements}
%\end{acknowledgements}


\begin{thebibliography}{}

%                          A&A references (bibitem, author limit)
%                          -------------------------------------    

\bibitem[]{}
Abramowicz M.A., Czerny B., Lasota J.-P., Szuszkiewicz E., 1988,
ApJ 332, 646

\bibitem[]{}
Abramowicz M.A., Chen X.-M., Granath M., Lasota J.-P., 1996, ApJ 471,
762

\bibitem[]{}
Angelini L., Parmar A.N., White N.E., 1997, Proc. IAU Colloquium 163,
Eds. D.T. Wickramasinghe, G.V. Bicknell, L. Ferrario, Port Douglas,
1996

\bibitem[]{}
Asai K., Dotani T. Nagase F., Mitsuda K., 2000, ApJS 131, 571

\bibitem[]{}
Ba\l uci\'nska-Church M., Church M.J., Oosterbroek T., et al.,
1999, A\&A 349, 495
%1323sax

\bibitem[]{}
Ba\l uci\'nska-Church M., Humphrey P.J., Church M.J., Parmar A.N.,
2000, A\&A 360, 583
%1624sax

\bibitem[]{}
Barret  D., Olive J.F., Boirin L., Done C., Skinner G.K., Grindlay J.E., 2000, ApJ 533. 329

\bibitem[]{}
Canizares C.R., Clark G.W., Li F.K., et al., 1975, ApJ 197, 457

\bibitem[]{}
Church M.J., 2001, Proceedings of 33rd Scientific Assembly of COSPAR, Warsaw,
July 2000, Adv Space Res in press

\bibitem[]{}
Church M.J., Ba\l uci\'nska-Church M., 2001, A\&A 369, 915
% Atoll sources

\bibitem[1995]{}
Church M.J., Ba\l uci\'nska-Church M., 1995, A\&A 300, 441
%1624exo

\bibitem[]{}
Church M.J., Inogamov N.A., Ba\l uci\'nska-Church M., 2001, A\&A in
preparation

\bibitem[1997]{}
Church M.J., Dotani T., Ba\l uci\'nska-Church M., et al., 1997, ApJ 491, 388
% 1916 asca

\bibitem[1998]{}
Church M.J., Ba\l uci\'nska-Church M., Dotani T., Asai K., 1998a, ApJ 504, 516
%0748

\bibitem[1998]{}
Church M.J., Parmar A.N., Ba\l uci\'nska-Church M., et al., 1998b, A\&A,
338, 556
% 1916 sax

\bibitem[]{}
Czerny B., Elvis M., 1987, ApJ 321, 305
% thin/thick disk transition: p_r/p_g

\bibitem[]{}
Ebisuzaki T., 1987, PASJ 39, 287

\bibitem[]{}
Ebisuzaki T., Hanawa T., Sugimoto D. 1984, PASJ 36, 551

\bibitem[]{}
Fujimoto M.Y., Hanawa T., Miyaji S., 1981, ApJ 246, 267

\bibitem[]{}
Frank J., King A.R., Raine D., 1992, ``Accretion Power in
Astrophysics'', Cambridge University Press
%H_eq formula

\bibitem[]{}
Hasinger G., Priehorsky W.C., Middleditch J., 1989,
ApJ 337, 843
%1st use of "Z-track"

\bibitem[]{}
Hasinger G., van der Klis M., 1989, A\&A 225, 79
%1st use of "Atoll"

\bibitem[]{}
Hirano T., Hayakawa S., Nagase F., Masai K., Mitsuda K., 1987, PASJ 39, 619

\bibitem[]{}
Hoffman J.A., Lewin W.H.G., Doty J., 1977, ApJ 217, L23

\bibitem[]{}
Igumenschev I.V., Abramowicz M.A., Novikov I.D., 1998, MNRAS 298, 1069

\bibitem[]{}
Inogamov N.A., Sunyaev R.A., 1999, Astron Lett 25, 269

\bibitem[]{}
Inoue H., Waki I., Koyama K., Matsuoka M., Ohashi T., Tanaka Y.,
Tsunemi H., 1984, PASJ 36, 831

\bibitem[]{}
Janiuk A., Czerny B., Siemiginowska A., 2000, ApJ 542, L33

\bibitem[]{}
Jones M.H., Watson M.G., 1989, Proc. of 23rd ESLAB Symposium, Bologna
%1624exo br+br

\bibitem[]{}
Joss P.C., Rappaport S.A., 1984, Ann Rev Astr Ap 22, 537

\bibitem[]{}
London R.A., Taam R.E., Howard W.M., 1984, ApJ 287, L27

\bibitem[]{}
London R.A., Taam R.E., Howard W.M., 1986, ApJ 306, 170

\bibitem[]{}
Madej J., 1991, ApJ 376, 161

\bibitem[]{}
Paczy\'nski B., 1983, ApJ 267, 315

\bibitem[]{}
Popham R., Sunyaev R.A., 2001, ApJ 547, 355

\bibitem[]{}
Schulz N.S., Hasinger G., Tr\" umper J., 1989, A\&A 225, 48
%Exo survey of LMXB: color-color, 2-cpt model

\bibitem[]{}
Shakura N.I., Sunyaev R.A., 1976, MNRAS 175, 613
%Thin disk

\bibitem[]{}
Smale A.P., Church M.J., Ba\l uci\'nska-Church M., 2001, ApJ 550, 962

\bibitem[]{}
Swank J.H., 2001, Proc. of ``Cosmic Explosions'', Maryland, October 2000,
Eds.  S.S. Holt \& W.W. Zhang, AIP Conf. Proc 522, in press

\bibitem[]{}
Swank J.H., Becker R.H., Boldt E.A., Holt S.S., Pravdo S.H.,
Serlemitsos P.J., 1977, ApJ 212, l73

\bibitem[]{}
Sztajno M., van Paradijs J., Lewin W.H.G., Tr\"umper J., Stollman G.,
Pietsch W., van der Klis M., 1985, ApJ 299, 487

\bibitem[]{}
van Paradijs J., 1984, Proc. of US-Japan Conf. on Galactic and Extragalactic
X-ray Sources, Tokyo 1984, Ed. Y. Tanaka

\bibitem[]{}
Watson M.G., Willingale R., King A.R., Grindlay J.E., Halpern J.,
1985, IAU Circ. 4051
%1624exo

\bibitem[1982]{}
White N.E., Swank J.H., 1982, ApJ 253, L61
%bulge

\bibitem[]{}
White N.E., Peacock A., Hasinger G., Masom K.O., Manzo G., Taylor B.G., 
Branduardi-Raymont G. 1986, MNRAS 218, 219

\bibitem[]{}
White N.E., Peacock A., Taylor B.G., 1985, ApJ 296, 475

\bibitem[]{}
Yoshida K., Mitsuda K., Ebisawa K., Ueda Y., Fujimoto R., Yaqoob T.
Done C., 1993, PASJ 45, 605

\bibitem[]{}
Zel'dovich Ya.B., Shakura N.I., 1969, Astron Zhur 46, 225 

\end{thebibliography}
\end{document}